\documentclass[conference]{IEEEtran}
\IEEEoverridecommandlockouts
\usepackage{amsmath,amssymb,amsfonts}
\usepackage{algorithmic}
\usepackage{graphicx}
\usepackage{comment}
\usepackage{textcomp}
\usepackage{enumitem}
\usepackage{xcolor}
\usepackage{tikz}
\usepackage{subcaption}  % For subfigures
\usepackage{graphicx}    % For images
\usepackage{caption}
\usepackage{xcolor, colortbl} % Load color and table coloring packages
\definecolor{customcyan}{HTML}{B0FCFF}
\def\BibTeX{{\rm B\kern-.05em{\sc i\kern-.025em b}\kern-.08em
    T\kern-.1667em\lower.7ex\hbox{E}\kern-.125emX}}
\begin{document}

%Exploiting D-Wave’s Quantum Annealing for Solving Multi-Objective Currency Arbitrage Problems*

\title{Solving Currency Arbitrage Problems using D-Wave Advantage2 Quantum Annealer}

\author{\IEEEauthorblockN{Lorenzo Mazzei}
\IEEEauthorblockA{\textit{DII, Univ. of Pisa} \\
L.go L. Lazzarino 1\\
56122 Pisa, Italy \\
l.mazzei7@studenti.unipi.it
}
\and
\IEEEauthorblockN{Giada Beccari}
\IEEEauthorblockA{\textit{DII, Univ. of Pisa}\\
L.go L. Lazzarino 1\\
56122 Pisa, Italy \\
g.beccari3@studenti.unipi.it
}
\and
\IEEEauthorblockN{Mirko Laruina}
\IEEEauthorblockA{\textit{DII, Univ. of Pisa} \\
L.go L. Lazzarino 1\\
56122 Pisa, Italy \\
m.laruina@studenti.unipi.it
}
\and
\hspace{8cm}\IEEEauthorblockN{Marco Cococcioni}
\IEEEauthorblockA{\hspace{8cm}\textit{DII, Univ. of Pisa} \\
\hspace{8cm}L.go L. Lazzarino 1\\
\hspace{8cm}56122 Pisa, Italy \\
\hspace{8cm}marco.cococcioni@unipi.it
}
}
\maketitle

%\hspace{2.5cm}\IEEEauthorblockN{Sebastian Feld}
%\IEEEauthorblockA{\hspace{2.5cm}\textit{Department of Quantum \& Computer Engineering} \\
%\hspace{2.5cm} Mekelweg 5,\\
%\hspace{2.5cm} 2628 CD, Delft, The Nederland \\
%\hspace{2.5cm}s.feld@tudelft.nl
%}
%\and

\begin{abstract}
 Quantum annealing has emerged as a powerful tool for solving combinatorial optimization problems efficiently, making use of the principles of quantum mechanics. Companies are increasingly investing in the market of quantum computers, providing the users with the possibility to solve these optimization problems by resorting to quantum computers. This paper explores how Quantum Annealing can be applied to the Currency Arbitrage (CA) optimization problem and its comparative performance against classical methods. 
 A key contribution of the work is an original formulation of the CA problem as a QUBO (Quadratic Unconstrained Boolean Optimization) problem. We test the speed of D-wave quantum annealer, using the recently released latest version (Advantage 2).
  %Our work also focuses on the performances of the prototypes of QPUs that D-Wave has been releasing, showing how the quantum approach could become the preferable one in terms of execution times.
\end{abstract}

\section{Introduction}

Tackling and solving an optimization problem is a task that can be carried out using multiple algorithms, each of them having their unique strengths and weaknesses. One of them, Simulated Annealing, aims at approximating the global optimum in a large search space. Simulated Annealing is still of interest, because it is one of the most promising concepts to give quantum superiority in Quantum Computing with its corresponding algorithm: Quantum Annealing \cite{D-Wave-QuantumAnnealing, Quinton2025, PhysRevE.58.5355, vesely2023findingoptimalcurrencycomposition}.
Quantum Annealing is a quantum computing technique designed to solve complex optimization problems \cite{Tasseff2024} by exploiting quantum mechanical phenomena. Unlike classical algorithms, Quantum Annealing leverages on quantum superposition and quantum tunneling to explore the vast search space, thereby speeding up the search for optimal solutions. One compelling application of Quantum Annealing is in financial optimization, particularly in solving Currency Arbitrage problems \cite{math12091291, Feld-article}. Currency Arbitrage involves exploiting discrepancies in exchange rates across different currencies to generate profits. 
Given the dynamic nature of global financial markets, the ability of Quantum Annealing to process vast amounts of interconnected data and detect optimal and near-optimal solutions in small time windows makes it a promising tool for financial institutions. This work explores the theoretical concepts behind Quantum Annealing as well as the formalization of the proposed Currency Arbitrage problem into the Quadratic Unconstrained Binary Optimization (QUBO) model \cite{glover2019tutorialformulatingusingqubo, D-Wave-QUBO}, which is the mathematical model to follow when the problem has to be approached with Quantum Annealing. The algorithm based on Quantum Annealing used in this paper is the one implemented by company D-Wave Systems. The performances of D-Wave's algorithms based on Quantum Annealing are compared to the ones of classical methods, both from the point of view of being able to reach the optimum as well as execution times. 
The paper is organized as follows. Section \ref{sec:problem} provides an introduction to the formulation of the Currency Arbitrage problem and how it can be expressed to be solved with Quantum Annealing. Section \ref{sec:comparison} includes the comparisons between the classical algorithm and Quantum Annealing. The analysis of the Advantage2 prototype2.6 QPU developed by D-Wave is carried out in Section \ref{sec:prototypes}. Conclusions are drawn in Section \ref{se:conclusion}.

\section{Problem Formulation}
\label{sec:problem}

The optimization problem we are trying to solve is the Currency Arbitrage problem. We want to capture trading loops of currencies that would allow us to make a profit. The goal is to start from a given currency, convert the money into other currencies, then close the loop by returning to the same currency we started with: if at the end of the loop we have more money than the quantity we started with, we made a profit and we can denote the loop as a ``profitable loop'' (see Fig. 1).
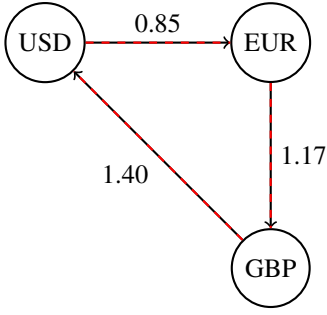
\begin{figure}[h]
    \centering
    \begin{tikzpicture}[node distance=3cm, auto, scale=1, transform shape]
        % Define nodes
        \node (USD) [circle, draw, thick] {USD};
        \node (EUR) [circle, draw, thick, right of=USD] {EUR};
        \node (GBP) [circle, draw, thick, below of=EUR] {GBP};

        % Define edges with exchange rates
        \path [->, thick] (USD) edge node {0.85} (EUR);
        \path [->, thick] (EUR) edge node {1.17} (GBP);
        \path [->, thick] (GBP) edge node {1.40} (USD);

        % Highlight arbitrage cycle
        \draw[thick, dashed, red] (USD) -- (EUR);
        \draw[thick, dashed, red] (EUR) -- (GBP);
        \draw[thick, dashed, red] (GBP) -- (USD);
    \end{tikzpicture}
    \caption{Currency Arbitrage Loop: The exchange rates allow for a profitable arbitrage opportunity by cycling through \newline USD → EUR → GBP → USD.}
    \label{fig:currency_arbitrage}
\end{figure}
\newline
%It is referred as ``loop'' because we start from a certain currency, we exchange with other currencies according to their exchange rates, and then we finish by returning to that same currency.

\subsection{QUBO Formulation}

A Quadratic Unconstrained Binary Optimization (QUBO) problem is defined using an upper-diagonal matrix $Q$, which is an $N \times N$ upper-triangular matrix of real weights, and $x$, a vector of binary variables of size $N$. The QUBO problem can be expressed concisely as follows:
\begin{equation}
    \label{eq:minxQx}
    \min_{x \in \{0, 1\}^N} x^T Q x
\end{equation}
More in detail, we are trying to minimize the following function:
\begin{equation}
    \label{eq:f(x)}
    f(x) = \sum_i Q_{i,i} x_i + \sum_{i<j} Q_{i,j} x_i x_j
\end{equation}
where the diagonal terms $Q_{i,i}$ are the linear coefficients and the nonzero off-diagonal terms $Q_{i,j}$ are the quadratic coefficients. To align our problem to the QUBO model formulation we formulate the problem with a set of variables \( x_{curr,pos} \), where \( curr \in \{1, \dots, N\} \) and \( pos \in \{1, \dots, K\} \). $K$ corresponds to the \textbf{loop length}, i.e. the number of currencies we want to have in the loop, while $N$ is the \textbf{total number of currencies}, i.e. how many currencies we consider when trying to find profitable loops of length $K$. To make the optimization problem non-trivial, we have to consider $K < N$.
\begin{equation}
    \label{eq:xik}
    x_{curr,pos} = 
    \begin{cases} 
        1 & \text{if currency } curr \text{ is present at position } pos \\ 
        0 & \text{otherwise}
    \end{cases}
\end{equation}
A loop is profitable if it brings a profit. This is true if its profitability factor is larger than $1$. The profitability factor is computed as the product of all the conversion rates of the currencies in the loop. We can denote the profitability factor $P$ of a loop $L$ as $P(L)$ and compute it as follows:
\begin{equation}
\label{eq:exchangeratesmultiplication}
P(L) = \prod_{k=1}^{K-1} ExchangeRates\left(x_{ik},x_{i(k+1)}\right)
\end{equation}
where $k$ is an index that corresponds to the position in the loop (e.g., in the Loop $[USD,EUR,GBP,USD]$ the USD currency is present in position $k=0$ and $k=3$, EUR in $k=1$, GBP in $k=2$). Using $P(L)$ expressed as a multiplication of several terms would raise the complexity of the problem, making it more than quadratic in complexity: this issue is solved by using the logarithm operator, which allows us to express the product as a sum by considering the logarithmic value of the exchange rate between any two currencies. We define it as: 
\begin{equation}
\label{eq:wij}
logRate_{ij} = - log( ExchangeRate_{k,k+1})
\end{equation}
Therefore, $P(L)$ is substituted with computing the sum of the \(logRate_{ij}\) terms for each consecutive couple of currencies $k$ and ${k + 1}$ in the loop $L$. This procedure reduces the complexity, allowing us to use a QUBO model to build the Hamiltonian that expresses the objectives and constraints of the problem. We also consider the inverse of the logarithm, i.e. the negative value of each \(logRate_{ij}\) term, to convert this optimization problem into a minimization one.

\subsection{Hamiltonian}
The Hamiltonian is the operator that represents the total energy of a quantum system. The system's evolution is governed by the Hamiltonian, which is specifically designed to guide the system toward the optimal solution of the optimization problem. The ground state (i.e., the lowest energy state) of the final Hamiltonian corresponds to the optimal solution.
To build the Hamiltonian for the Currency Arbitrage problem we want to solve, the objective(s) and the constraints have to be outlined: we can map each objective and constraint into their own Hamiltonian.
\newline
\textbf{Objectives}:
    \begin{enumerate}
        \item \(\mathcal{H}_A\): the sum of the $logRate_{ij}$ terms should be minimized in order to obtain the most profitable loop.
        \item \(\mathcal{H}_B\): a currency appearing in consecutive positions in the loop should be rewarded, to facilitate shorter loops.
    \end{enumerate}
\textbf{Constraints}:
    \begin{itemize}
        \item \(\mathcal{H}_C\): a position in the loop cannot contain more than one currency.
        \item \(\mathcal{H}_D\): the loop must begin and end with the same currency.
        \item \(\mathcal{H}_E\): no position of the loop must be left empty.
    \end{itemize}
These Hamiltonians are formulated in the following way:
\begin{equation}
    \label{eq:HA}
    \mathcal{H}_A = A \sum_{i=1}^{N} \sum_{j=1}^{i-1} logRate_{ij} \sum_{k=1}^{K-1} x_{ik} x_{j(k+1)}
\end{equation}

\begin{equation}
    \label{eq:HB}
    \mathcal{H}_B = B \sum_{k=1}^{K} \sum_{i=2}^{N} x_{ik} \sum_{j=1}^{i-1} x_{jk}
\end{equation}

\begin{equation}
    \label{eq:HC}
    \mathcal{H}_C = C \sum_{i=1}^{N} \left( 1 - (x_{i1} - x_{iK})^2 \right)
\end{equation}

\begin{equation}
    \label{eq:HD}
    \mathcal{H}_D = D \sum_{i=1}^{N} \sum_{k=1}^{K-1} x_{ik} x_{i\left(k+1\right)}
\end{equation}

\begin{equation}
    \label{eq:HE}
    \mathcal{H}_E = E \sum_{k=1}^{K} \sum_{i=1}^{N} x_{ik}
\end{equation}
\begin{figure}[!hbtp]
    \centering    \includegraphics[width=1\linewidth]{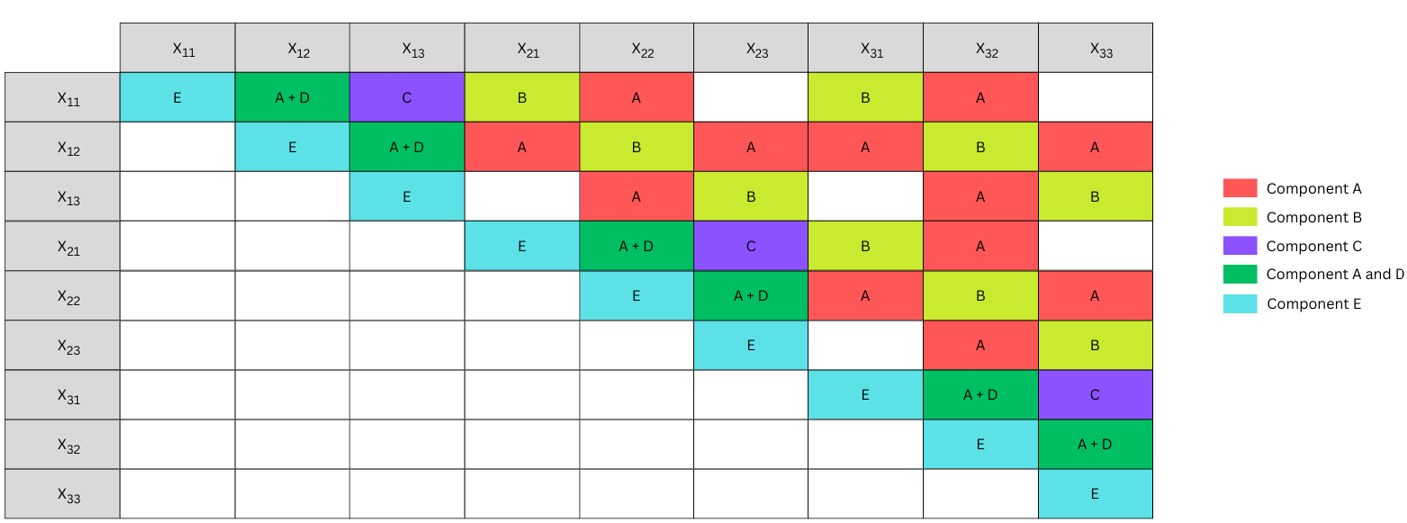}    \caption{\label{fig:QUBO_Matrix}QUBO Matrix.}
\end{figure}
Where $A,B,C,D,E$ are the weights of the Hamiltonians. Assigning an appropriate value to each of the weights is necessary to obtain satisfying results: e.g., the larger the value given to weight $A$ of \(\mathcal{H}_A\), the more we are pushing the algorithms to solve the problem in search of more profitable loops, but this could lead to solutions that are not feasible by violating one or more of the constraints. This means that there is a trade-off between solving the problems for more profitable loops and obtaining feasible solutions that respect the constraints imposed. \newline
Therefore the total Hamiltonian that includes the objectives and constraints of our problem is obtained as follows:
\begin{equation}
    \label{eq:hamiltoniansums}
    \mathcal{H} = \mathcal{H}_A + \mathcal{H}_B + \mathcal{H}_C + \mathcal{H}_D +\mathcal{H}_E 
\end{equation}    

We can now define the QUBO matrix, which is a square matrix that represents a QUBO problem according to the formulation reported in Eq. \eqref{eq:f(x)}. In our case, the coefficients correspond to the weights of the Hamiltonian and $logRate_{ij}$; the binary variables correspond to the ones defined in Eq. \eqref{eq:xik}.
For the Hamiltonian $\mathcal{H}_E$, the interactions are only between the same variables, which means that the weights will represent the diagonal elements of the QUBO matrix.
In all the other Hamiltonians the interactions are between different variables, so each of their coefficients are off-diagonal elements. Since the problem regards exchange rates, the matrix is a triangular matrix, therefore we can analyze only the upper or lower triangular part of it. Graphically, we consider the upper part. Let us take as an example a Currency Arbitrage problem with three currencies, the QUBO matrix with all the coefficients included is reported in Fig. \ref{fig:QUBO_Matrix}: 
\begin{itemize}
    \item $\mathcal{H}_A$: the coefficient given by $A \cdot logRate_{ij}$ contributes to the Hamiltonian only if currency $i$ is in position $k$ in the loop and currency $j$ is in position $k+1$, therefore it fills the cells where the currency $j$ is one position ahead of the currency $i$.
    \item $\mathcal{H}_B$: the coefficient given by $B$ contributes to the Hamiltonian only if currency $j$ is not the same currency as $i$ for the same position $k$.
    \item $\mathcal{H}_C$: the coefficient given by $C$ contributes to the Hamiltonian only if currency $i$ is the same as currency $j$ in position $1$ and in the last position of the loop.
    \item $\mathcal{H}_D$: the coefficient given by $D$ contributes to the Hamiltonian only if currency $i$ is the same as currency $j$ in position $k$ and in position $k+1$.
    \item $\mathcal{H}_E$: the coefficient given by $E$ represents the diagonal elements as mentioned before.
\end{itemize}
 
\section{Comparison between Classical Solvers and Quantum Annealing}
\label{sec:comparison}
An interesting behavior examined in this work was how quickly the algorithms used were able to find the optimal solution: in Fig. 3 we reported the total execution times of the algorithms, more specifically how much time it took each algorithm to compute all the reads they were set up for (through the \textit{num\_reads} parameter). Therefore, these data highlight only the speed of each algorithm to execute a given number of reads. The goal of this work though is to find optimal solutions for the Currency Arbitrage problems as quickly as possible, so it would be more useful to gather insights on how many reads each algorithm actually need to find the optimal value: for instance, if the DWaveSampler algorithm was set up to execute a total of $1,000$ reads, but it was observed that it finds the optimum value with fewer than $50$ reads for the majority of the trials, then the speed performance of the DwaveSampler algorithm should be measured according to that $50$ reads, i.e. the actual reads it takes the algorithm to get to the optimal value. These line of process involves both the classical algorithms and the DwaveSampler algorithm based on Quantum Annealing. The classical algorithms taken into account for this comparison are \textit{SimulatedAnnealingSampler} and \textit{TabuSampler}, with the latter being based on Tabu Search, a metaheuristic algorithm that solves optimization problems by exploring the solution space while avoiding cycles and revisiting previously explored solutions by making use of a memory structure called \emph{tabu list}. For Quantum Annealing we used \textit{D-WaveSampler}, a sampler that uses a D-Wave quantum computer to sample from the QUBO formulation of our problem (which is more specifically a Binary Quadratic Model), leveraging on the Quantum Annealing algorithm to find the optimal or near-optimal solutions \cite{D-Wave-DWaveSampler}.\newline
For this purpose, tests were run to check the number of reads that were necessary to each algorithm to reach the optimal solution for the first time. We used D-Wave's \textit{ExactSolver}, a sampler that performs brute-force enumeration over all possible states of a Binary Quadratic Model, to retrieve the optimal solution for the configuration of currencies considered so that it could be used as the ground truth for the other algorithms; while running the algorithms it is not possible to stop their execution when the optimal solution is reached, because there is no way to control the flow of executions of the algorithms until they have finished their number of reads, so the workaround method used in this work was to save the samples obtained by the algorithms in the exact order they were found and then iterate all these results. For each algorithm, once the sample representing the optimal solution was found, the index associated with that sample was retrieved. At the end of this process we were able to obtain the first read of each algorithm when the optimal value was found. \newline
\begin{figure*}[!htbp]
    \begin{minipage}[b]{0.49\textwidth}
        \centering        \includegraphics[width=\linewidth]{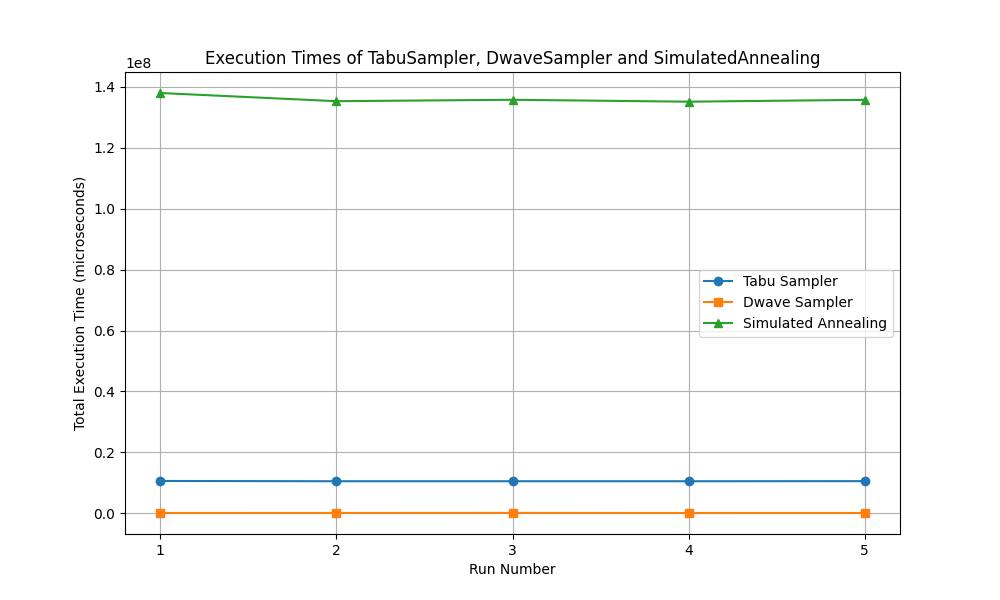}
        \caption{Total execution times} \label{fig:execution_times_of_the_3_solvers}
    \end{minipage}
    \hfill
    \begin{minipage}[b]{0.49\textwidth}
        \centering        \includegraphics[width=\linewidth]{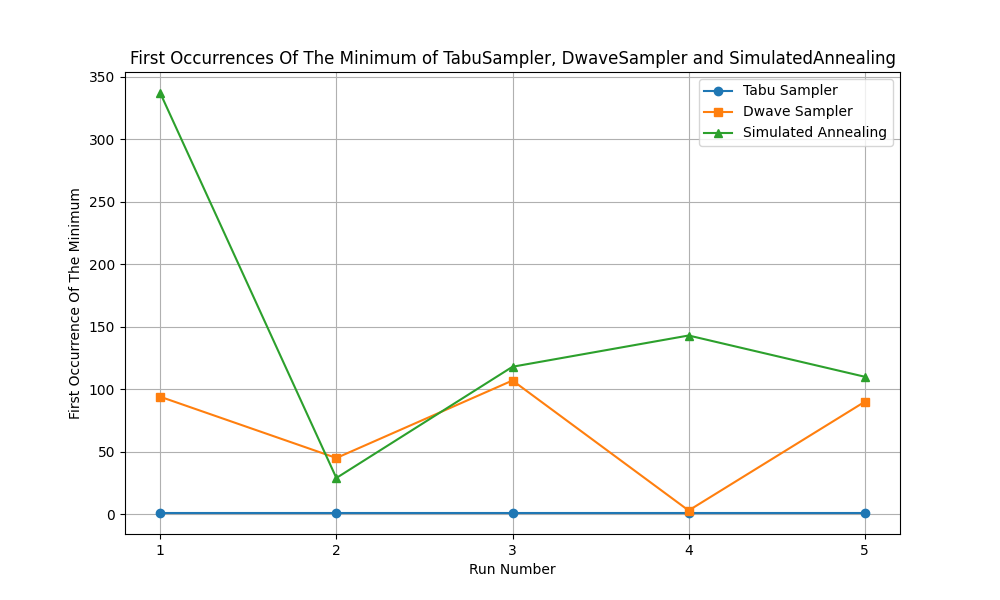}
        \caption{First reads needed to find the optimum} \label{fig:first_minimum_occurrences_of_the_3_solvers}
    \end{minipage}
\end{figure*}
\begin{figure}[!hbtp]
    \centering    \includegraphics[width=1\linewidth]{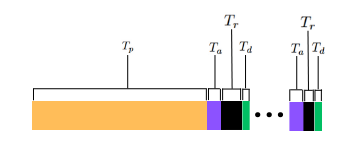}    \caption{\label{fig:QPU_access_time}Components of the QPU access time.}
\end{figure}
The results are reported in Fig. \ref{fig:first_minimum_occurrences_of_the_3_solvers}, where 'Run Number' on the x-axis refers to the corresponding batch of 500 reads. From that, it can be inferred that Simulated Annealing is on average the algorithm that requires the higher number of reads to reach the optimum for the first time; D-WaveSampler takes fewer reads to get the optimum solution; TabuSampler is the best in this sense since it requires just one read; this can be due to the fact that if the problem size is small or the structure of the solution space is such that TabuSampler's heuristic rules can efficiently guide it to the optimal solution, a single read can identify the optimal answer. \newline
From Fig. \ref{fig:execution_times_of_the_3_solvers} it can be observed that Simulated Annealing is also the slowest between the algorithms, which confirms that D-WaveSampler is the fastest of the three in the total execution time considering all the reads, but as seen in Fig. \ref{fig:first_minimum_occurrences_of_the_3_solvers} it needs more reads than TabuSampler. Therefore, it proved interesting to compute an estimate on the actual time needed by TabuSampler and D-WaveSampler to reach the read corresponding to the optimum solution without considering Simulated Annealing in the process since it would come out as the slower of the three apriori. Since, as previously mentioned, there is no way of retrieving information during the execution of the algorithms, estimations had to be made. 
TabuSampler reached the optimum in the first read all the times in this configuration and it took around $24,000$ microseconds with one read. For D-WaveSampler, the way its execution time is computed required a more detailed study of the time components: the execution time reported in the figures refers to the \textit{qpu\_access\_time} parameter. The time provided by this parameter is actually composed of multiple entities, all of them being reported in the D-Wave Documentation which also provides the following image (more details can be found in \cite{D-Wave-Timing}) to display them visually:
Overall, the \textit{qpu\_access\_time} can be computed as follows:

\begin{equation}
    \label{eq:Tqpu}
    T_{qpu} = T_p + \Delta + T_s
\end{equation}

\begin{table}[htbp]
\scriptsize
    \caption{Comparison of QPU metrics for different \texttt{\textit{num\_reads}} values.}
    \begin{center}
        \begin{tabular}{|c|c|c|c|c|}
            \hline
            \textbf{Metric} & \multicolumn{4}{|c|}{\textbf{\texttt{\textit{num\_reads}}}} \\
            \cline{2-5}
            \textbf{Name} & \textbf{\textit{1}} & \textbf{\textit{10}} & \textbf{\textit{100}} & \textbf{\textit{500}} \\
            \hline            \texttt{qpu\_access\_time} & 15900 & 17133 & 34145 & 91141 \\
            \hline            \texttt{qpu\_sampling\_time} & 117 & 1370 & 18384 & 75380 \\
            \hline            \texttt{qpu\_programming\_time} & 15782 & 15762 & 15761 & 15761 \\
            \hline            \texttt{qpu\_readout\_time\_per\_sample} & 47 & 66 & 113 & 80 \\
            \hline        \texttt{qpu\_delay\_time\_per\_sample} & 20 & 20 & 20 & 20 \\
            \hline
        \end{tabular}
        \label{tab:comparison}
    \end{center}
\end{table}
$T_p$ (programming time) is a one-time initialization step to program the QPU. $T_s$ represents the time for all the samples. \(\Delta\) is denoted as \textit{qpu\_access\_overhead\_time} and it is an initialization time spent in low-level operations, roughly 10-20 ms for Advantage systems. In the D-Wave documentation it is stated that \(\Delta\) is not included in the \textit{qpu\_access\_time} field. Since \(\Delta\) is still a time component that has a weight in the total execution time, we analyzed the results with it included: if we consider \(\Delta\) to be always 20ms (the worst case scenario according to the documentation) and add it to the total execution times of D-WaveSampler, the relative results would actually be the same. Indeed, even if we consider the worst time of D-WaveSampler between all the ones obtained during the trials, which was 107,482\(\mu\)\text{s}, and add to it the 20ms given by \(\Delta\), we obtain 127,482\(\mu\)\text{s} which is still almost two orders of magnitude smaller than the total execution times of TabuSampler (see Fig. \ref{fig:execution_times_of_the_3_solvers}). \textit{Ts} can be estimated in the following way: the time for a single sample is given by the sum of the \textit{anneal time} $T_a$ (the parameter set by the user when executing D-WaveSampler), \textit{readout time} $T_r$ (time required to read the sample from the QPU) and \textit{thermalization time} $T_d$ (time required by the QPU to regain its initial temperature). 
Therefore, if we multiply the time for a single sample with the \textit{total number of reads} $R$, we obtain $T_s$:
\begin{equation}
\label{eq:Ts}
T_s = R \cdot (T_a + T_r + T_d)
\end{equation}
From how $T_{qpu}$ is computed it can be deduced that the only component on which the developer has control is $T_a$. It also can be inferred that $T_p$ can account for most of the total time of D-WaveSampler if we use very few reads. \newline
Overall, the behavior of the time components of D-WaveSampler can be observed directly by accessing the \textit{info['timing']} field of the SampleSet object returned by the execution of D-WaveSampler. \newline
\begin{figure*}[!htbp]
    \centering
    \begin{minipage}[b]{0.2\textwidth}
        \centering        \includegraphics[width=\linewidth]{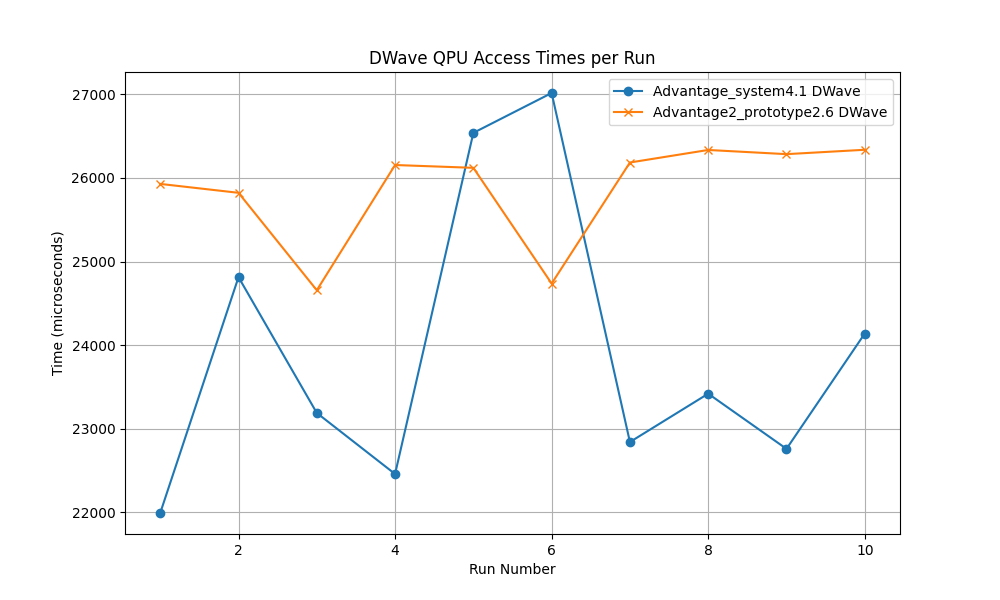}
        \caption{First run with 50 reads}    \label{fig:50reads_d-wave_qpu_access_times}
    \end{minipage}
    \hfill
    \begin{minipage}[b]{0.2\textwidth}
        \centering        \includegraphics[width=\linewidth]{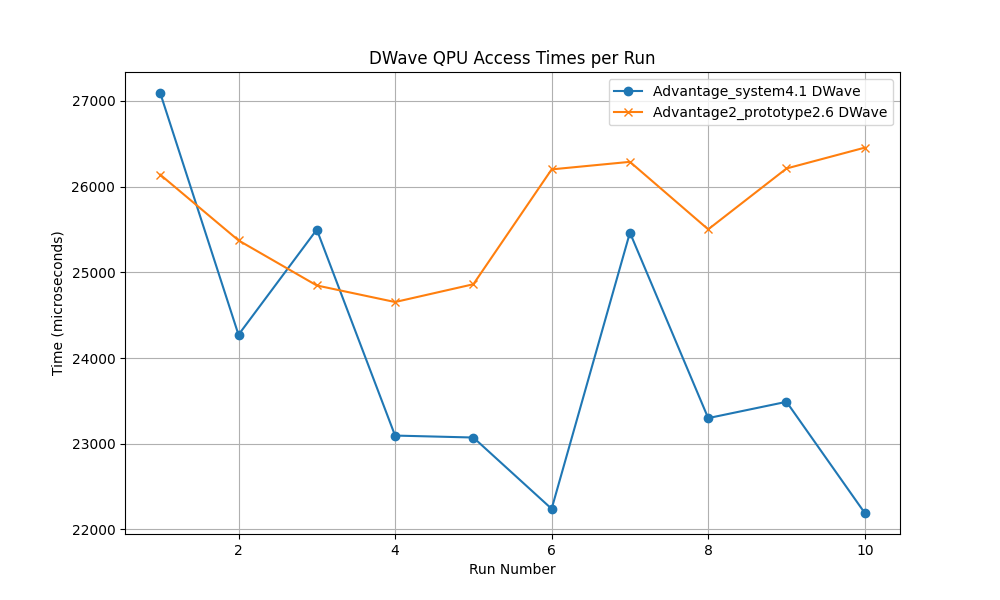}
        \caption{Second run with 50 reads} \label{fig:50reads_d-wave_qpu_access_times_second_run.png}
    \end{minipage}
    \hfill
    \begin{minipage}[b]{0.2\textwidth}
        \centering        \includegraphics[width=\linewidth]{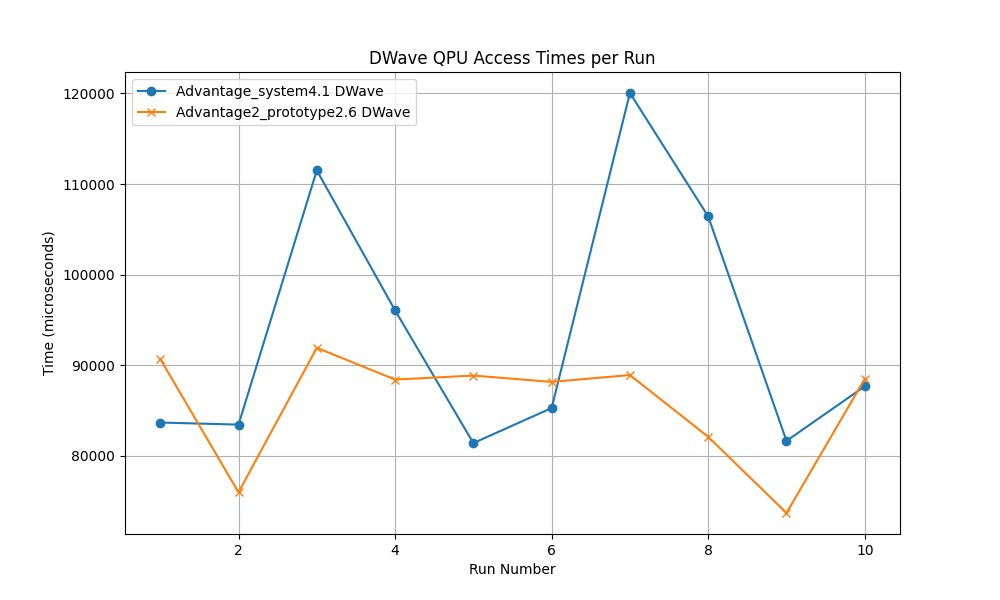}
        \caption{First run with 500 reads}    \label{fig:500reads_d-wave_qpu_access_times}
    \end{minipage}
    \hfill
    \begin{minipage}[b]{0.2\textwidth}
        \centering        \includegraphics[width=\linewidth]{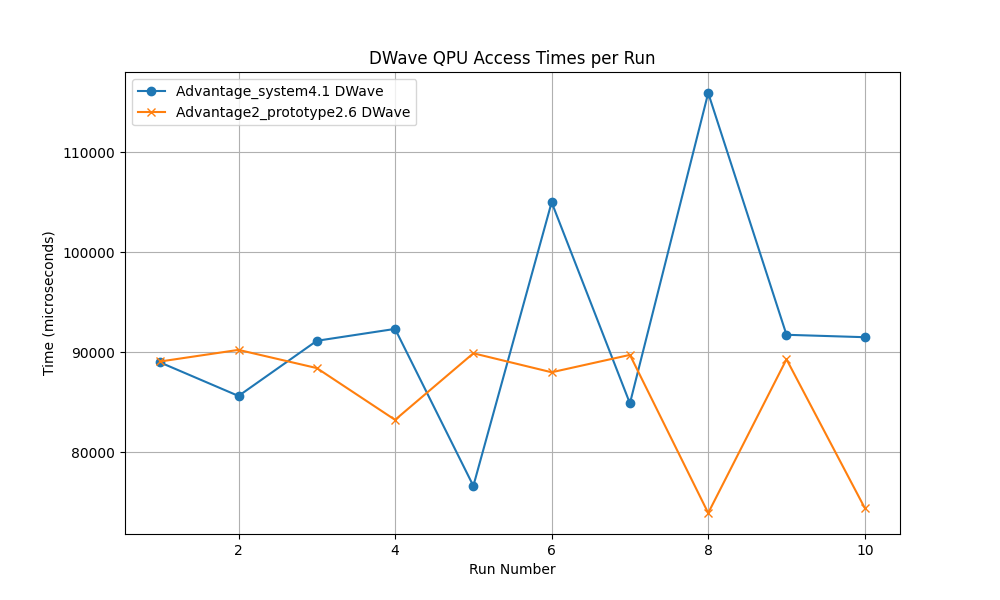}
        \caption{Second run with 500 reads} \label{fig:500reads_d-wave_qpu_access_times_second_run.png}
    \end{minipage}
\end{figure*}

\begin{figure*}[!htbp]
    \centering
    \begin{minipage}[b]{0.2\textwidth}
        \centering        \includegraphics[width=\linewidth]{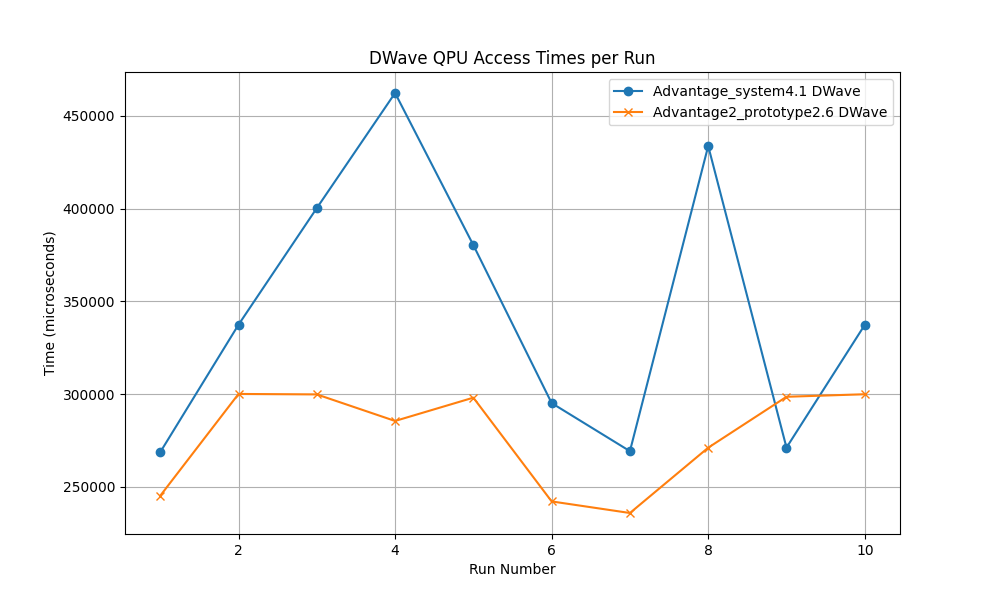}
        \caption{First run with 2000 reads}    \label{fig:2000reads_d-wave_qpu_access_times}
    \end{minipage}
    \hfill
    \begin{minipage}[b]{0.2\textwidth}
        \centering        \includegraphics[width=\linewidth]{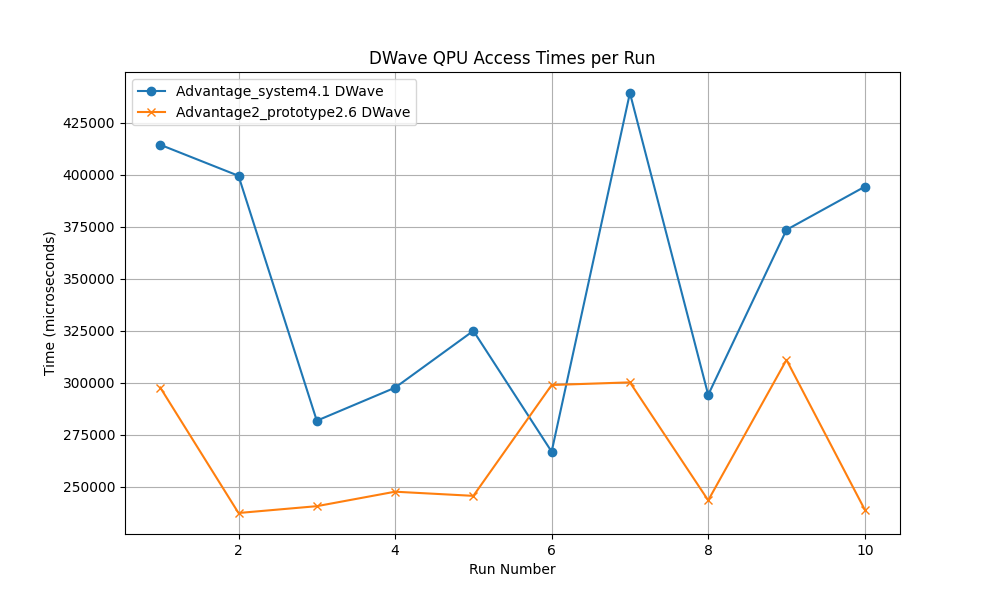}
        \caption{Second run with 2000 reads} \label{fig:2000reads_d-wave_qpu_access_times_second_run.png}
    \end{minipage}
    \hfill
    \begin{minipage}[b]{0.2\textwidth}
        \centering        \includegraphics[width=\linewidth]{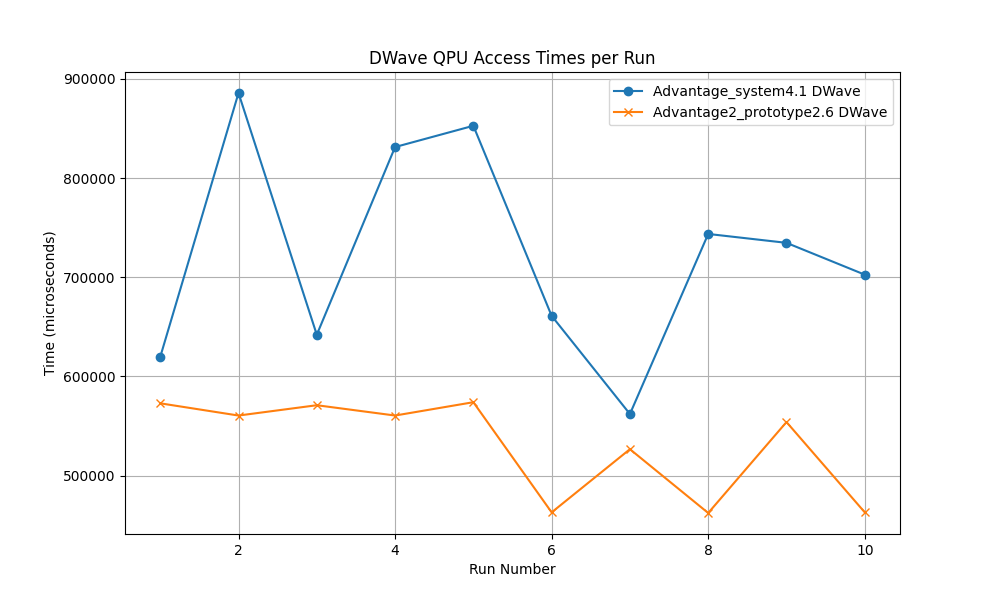}
        \caption{First run with 4000 reads}    \label{fig:4000reads_d-wave_qpu_access_times}
    \end{minipage}
    \hfill
    \begin{minipage}[b]{0.2\textwidth}
        \centering        \includegraphics[width=\linewidth]{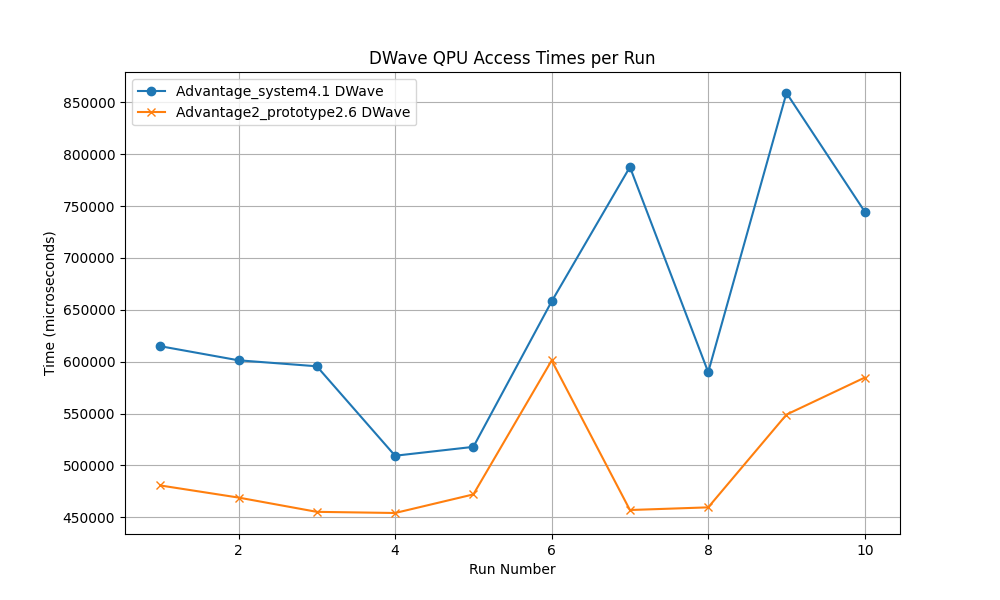}
        \caption{Second run with 4000 reads} \label{fig:4000reads_d-wave_qpu_access_times_second_run.png}
    \end{minipage}
\end{figure*}
The execution of D-WaveSampler considering \textit{anneal\_time} = 50 for different \textit{num\_reads} values brought the results reported in Table~\ref{tab:comparison} for the mentioned time components. From that, it can be observed that \textit{qpu\_delay\_time\_per\_sample} has a constant value independently from the \textit{num\_read} value; the same can be said for \textit{qpu\_programming\_time} with negligible variations; moreover, even running D-WaveSampler on different days, i.e., with different rates between the currencies, does not affect this parameter values, so it could be considered constant from what it was observed in this work;
\textit{qpu\_readout\_time\_per\_sample} instead does not follow a pattern in its values, it has a non-deterministic behavior.

Overall, it can be inferred that \textit{Simulated Annealing} is the slowest both in total time and in the time needed to reach the optimum for the first time; \textit{TabuSampler} reaches the optimum in the first read in the tests made and takes around 24,000\(\mu\)\text{s} if it does only one read; D-WaveSampler is the one that requires the shortest amount of time to perform a single read per se, but it requires few reads to reach the optimal value while also having to deal with overheads like \textit{qpu\_programming\_time}. TabuSampler is therefore, at the moment this work was carried out, the fastest in reaching the optimum, but D-WaveSampler is promising and could surpass it if the future hardware will bring lower overhead times. For this purpose, a comparison with a new QPU prototype that was being developed by D-Wave is explored in Sec. \ref{sec:prototypes} of this paper.
 
\section{Advantage2 Prototypes}
\label{sec:prototypes}
Analyses on the different D-Wave QPUs performances have already being carried out in other studies \cite{Willsch2022}. But currently D-Wave is developing a new QPU called Advantage2. They are creating prototypes of this new version and they update these prototypes as time passes. For instance, during August and September of 2024, the \textit{Advantage2\_prototype2.4} was being developed and made available to users. During the tests made in this work, \textit{Advantage2\_prototype2.4} was used to see if it allowed to have efficient results as with \textit{Advantage\_system4.1}, but with better times. The results obtained were promising and D-Wave was going to put out better prototypes as time passed on, so we considered this an aspect that we had to follow up with other tests. In October and November 2024 \textit{Advantage2\_prototype2.5} was created, then in November 2024 and December 2024 \textit{Advantage2\_prototype2.6} was created. To check if the results kept being promising, we performed tests on \textit{Advantage2\_prototype2.6}: in particular, D-WaveSampler was tried both with \textit{Advantage\_system4.1} and with \textit{Advantage2\_prototype2.6}, the solutions were then compared as well as their execution times. The results of these tests are reported in Fig. \ref{fig:50reads_d-wave_qpu_access_times},\ref{fig:50reads_d-wave_qpu_access_times_second_run.png},\ref{fig:500reads_d-wave_qpu_access_times},\ref{fig:500reads_d-wave_qpu_access_times_second_run.png},\ref{fig:2000reads_d-wave_qpu_access_times},\ref{fig:2000reads_d-wave_qpu_access_times_second_run.png},\ref{fig:4000reads_d-wave_qpu_access_times},\ref{fig:4000reads_d-wave_qpu_access_times_second_run.png}.
\newline
The parameter used to measure the execution times is the \textit{qpu\_access\_time} parameter, which represents the actual time spent on the QPU for solving the problem, as already mentioned in Sec. \ref{sec:comparison} of this paper. Time is noted in microseconds as confirmed in the D-Wave documentation at the voice 'SAPI Timing Fields'. These tests were made considering $N = 5$ and
$K = 4$. The \textit{qpu\_access\_time} was measured considering four different values for the \textit{\textit{num\_reads}} parameter: 50; 500; 2000; 4000. For each of these values, two batches of ten executions each were carried out: i.e. if we take the \textit{\textit{num\_reads}} = 50 case,  D-WaveSampler was ran ten different times for the first batch and then another ten times for the second batch; this process was repeated also for \textit{\textit{num\_reads}} = 500, \textit{num\_reads} = 2000 and \textit{num\_reads} = 4000. The goal was to examine if the difference in time executions between the two QPUs remained constant or if the gap actually grew when considering a growing amount of \textit{num\_reads}. The arithmetic means of the executions times of all the trials in Fig. \ref{fig:50reads_d-wave_qpu_access_times},\ref{fig:50reads_d-wave_qpu_access_times_second_run.png},\ref{fig:500reads_d-wave_qpu_access_times},\ref{fig:500reads_d-wave_qpu_access_times_second_run.png},\ref{fig:2000reads_d-wave_qpu_access_times},\ref{fig:2000reads_d-wave_qpu_access_times_second_run.png},\ref{fig:4000reads_d-wave_qpu_access_times},\ref{fig:4000reads_d-wave_qpu_access_times_second_run.png} are reported in Table \ref{tab:advantage_comparison}.\newline
\begin{table}[!hbtp]
    \caption{Comparison of Advantage System Metrics Across Two Batches}
    \begin{center}
        \begin{tabular}{|c|c|c|c|c|}
            \hline
            \textbf{} & \multicolumn{2}{|c|}{\textbf{Batch 1}} & \multicolumn{2}{|c|}{\textbf{Batch 2}} \\
            \cline{2-5}
            \textbf{\textit{num\_reads}} & \textbf{System4.1} & \textbf{Prototype2.6} & \textbf{System4.1} & \textbf{Prototype2.6} \\
            \hline
            \texttt{50} & 23917 & 25856 & 23971 & 25653 \\
            \hline
            \texttt{500} & 93725 & 85737 & 92374 & 85617 \\
            \hline
            \texttt{2000} & 345590 & 277681 & 348638 & 266169 \\
            \hline
            \texttt{4000} & 723505 & 530786 & 647858 & 498289 \\
            \hline
        \end{tabular}
        \label{tab:advantage_comparison}
    \end{center}
\end{table}
From that it can be observed that for $\textit{num\_reads} = 50$ \textit{Advantage system4.1} performs better in execution time, for both the batches. This does not hold true when \textit{num\_reads} = 500, where \textit{Advantage2 prototype2.6} already starts performing faster. It then can be seen that the difference in times grows wider when $\textit{num\_reads} = 2000$ and even more when $\textit{num\_reads} = 4000$. Overall, \textit{Advantage2 prototype2.6} already seems to perform faster than the system used for this work, therefore when the actual \textit{Advantage2} will be complete and released far better results are to be expected.

\section{Conclusion}
\label{se:conclusion}
In this paper we first provide an original QUBO formulation for the currency arbitrage problem and then we explored
the potential of Quantum Annealing in solving it, with an emphasis on evaluating its effectiveness with respect to classical counterparts (Simulated Annealing and Tabu Search). The APIs from D-Wave were used to access the Quantum Annealing based algorithm D-WaveSampler; we compared the obtained performance with the mentioned classical algorithms, both from the point of view of total execution times and from point of view of reaching the optimum for the first time. The results of tests made with a more recent prototype of D-Wave's QPU (Advantage 2) were also reported, highlighting how D-Wave's Quantum Annealing is becoming faster and faster, with the promise of outperforming the classical algorithms. 
Overall, D-wave's Quantum Annealer proved to be a valid alternative at solving our optimization problem, and it has the potential to become better than classical algorithms
in its future, improved, versions (more qubits, better topology, increased coherence time, etc.).
%and will still see support in the future, as at the current time D-Wave is still working towards a new QPU which could be interesting to expand upon when it will be fully released.

\section*{Acknowledgment}
We deeply thank Sebastian Feld for his help in the creation of the QUBO formulation of this problem.\\
This study was carried out within the ICSC National Centre on HPC, 
Big Data and Quantum Computing - SPOKE 10 (Quantum Computing) and received funding from the European 
Union Next-GenerationEU - National Recovery and Resilience Plan (NRRP) – MISSION 4 COMPONENT 2,
INVESTMENT N. 1.4 – CUP N. I53C22000690001. 
This work reflects solely the opinions and views of the authors, and neither the European Union nor the European Commission can be held responsible for them.

\bibliographystyle{IEEEtran}
%\bibliography{paperBibliography}
%\bibliography{bib}


\end{document}